\documentclass[prb,showpacs,floatfix,footinbib,twocolumn]{revtex4}
\usepackage{graphicx}
\usepackage{dcolumn}
\usepackage{epsfig}
\usepackage{float}
\usepackage{bm}
\usepackage{bibentry}

\begin{document}

\bibliographystyle{apsrev}

\title{Infrared Hall conductivity of Na$_{0.7}$CoO$_2$}
\author{E. J. Choi$^{1}$, S. H. Jung$^{1}$, J. H. Noh$^{1}$, A. Zimmers$^{2}$, D. Schmadel$^{2}$ H. D. Drew$^{2}$,
J.Y. Son $^{3}$ and J.H. Cho $^{3}$}

\affiliation{$^1$Department of Physics, University of Seoul, Seoul
130-743, Republic of Korea} \affiliation{$^2$ Dept. of Physics,
Univ. of Maryland, College Park, MD20742, USA} \affiliation{$^3$
RCDAMP and Department of Physics, Pusan National University Pusan,
Republic of Korea}

%
%

\begin{abstract}
We report infrared Hall conductivity $\sigma_{xy}(\omega)$ of
Na$_{0.7}$CoO$_2$ thin films determined from Faraday rotation angle
$\theta_{F}$ measurements. $\sigma_{xy}(\omega)$ exhibits two types
of hole conduction, Drude and incoherent carriers. The coherent
Drude carrier shows a large renormalized mass and Fermi liquid-like
behavior of Hall scattering rate, $\gamma_{H} \sim aT^{2}$. The
spectral weight is suppressed and disappears at T = 120K. The
incoherent carrier response is centered at mid-IR frequency and
shifts to lower energy with increasing T. Infrared Hall constant is
positive and almost independent of temperature in sharp contrast
with the dc-Hall constant.

\end{abstract}


\maketitle

The sodium cobalt oxide Na$_{x}$CoO$_{2}$ has attracted much
attention recently since the discovery of large thermopower at x$>$
0.7 \cite{terasaki} and T=5K superconductivity at x=0.3.\cite{super}
The material has a layered crystal structure in which edge-sharing
CoO$_6$ octahedra form 2D plane separated by the charge-providing
Na$^{+2}$ spacer. The CoO$_{2}$ plane has triangular bonding
symmetry which brings about geometrical frustration of the Co spins.
Carrier transport in this conducting plane can potentially lead to
exotic phase such Anderson's resonant valence bond (RVB) state or
non-s wave superconducting pairing.\cite{baskaran} With varying Na
content, x , Na$_{x}$CoO$_{2}$ exhibits a rich phase diagram which
includes the superconductivity (x=0.35),\cite{super} charge ordered
insulating state (x=0.5),\cite{Foo} Curie-Weiss metal (x=0.7), and
spin-density wave state (x$>$0.65).\cite{sdw} The large thermopower
observed at high x also makes the material interesting for
applications .\cite{terasaki}

Na$_{x}$CoO$_{2}$ with x=0.7 is a host compound of the series and it
shows its own unusual properties at low temperature. Dc resistivity
shows linear-T behavior at T $\leq $ 100K. The quasiparticles seen
in angle resolved photoemission (ARPES) experiment decreases in
intensity with T and disappears at T $\geq $ 100K.\cite{Hasan04}
Also, in spite of moderate metallicity ($\rho$ = 0.1m$\Omega$ at
30K), the magnetic susceptibility data is not Pauli-like but rather
the Co spins show local magnetic moments. The Hall coefficient
increases with temperature without saturation up to as high as
T=500K.\cite{wang} The positive Hall sign changes to negative at
T$<$ 200K. These anomalous properties suggest that Na$_{x}$CoO$_{2}$
is another example of a strongly correlated electron system.

In Na$_{0.7}$CoO$_{2}$, the active Co t$_{2g}$ bands are split by
crystal field into e$_{g}$ and d$_{z^{2}}$  bands. The e$_{g}$ band
is fully occupied. The electronically active d$_{z^{2}}$  band is
half filled (occupied by 1 electron per Co) at x=0 and adding
electrons at x$>$0 can lead to a doped Mott-Hubbard system where the
fundamental physics is exotic and not well understood and therefore
it is currently subject of intense study.\cite{Imada}

In this paper, we report infrared Hall effect of Na$_{0.7}$CoO$_{2}$
from Faraday angle measurements on thin film samples. The experiment
probes the Hall conductivity $\sigma_{xy}$ at an infrared frequency.
The study of $\sigma_{xy}$ provides invaluable informations on
charge dynamics of strongly correlated metal, often complimentary to
the conventional optical conductivity $\sigma_{xx}$ as successfully
demonstrated in high-Tc superconductor YBaCuO,\cite{htsc1,htsc1a}
Bi$_{2}$Sr$_{2}$CaCu$_{2}$O$_{8}$,\cite{htsc2} and electron-doped
Pr$_{2-x}$Ce$_{x}$CuO$_{4}$.\cite{htsc3}

Epitaxial $\gamma$-Na$_{0.7}$CoO$_{2}$
 thin films (1000{\AA} thick) were grown by
pulsed laser deposition technique on SrTiO$_{3}$ substrates. The
films have (0001)orientation where the CoO$_{2}$ hexagonal plane is
parallel with the substrate plane. Details of the sample growth and
characterization was published elsewhere.\cite{Cho} We measured
dc-resistance and dc-Hall effect of the films for 4K $\leq $ T $\leq
$ 350K range. Both results showed similar temperature dependent
results as those for single crystal Na$_{0.7}$CoO$_{2}$ , suggesting
high quality of the films. Faraday rotation angle of the film was
measured using polarized infrared CO$_{2}$ laser with frequency
$\omega$ = 1100 cm$^{-1}$  and a photoelastic modulator at 30K $\leq
$ T $\leq $ 300K. The substrate contribution was independently
measured and subtracted from data. The details of measurement
technique are described elsewhere. \cite{Cerne}

%
%

\begin{figure}[t]
\vspace*{-0.2cm}\centerline{\includegraphics[width=3.5in,angle=0]{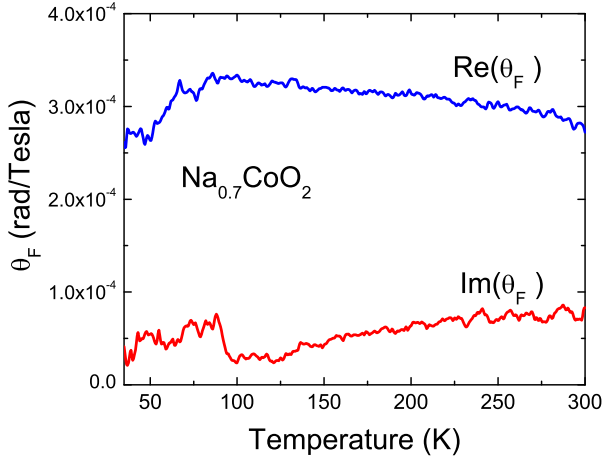}}
\vspace*{0.0cm}%
%
\caption{Complex Faraday rotation angle $\theta_{F}$ of
Na$_{0.7}$CoO$_2$ film measured at infrared frequency $\omega$ =
1100 cm$^{-1}$.} \label{fig:IRhallfig1}
\end{figure}
Figure 1 shows the infrared complex Faraday angle $\theta_{F}$ (real
and imaginary part) of a Na$_{0.7}$CoO$_2$ film at $\omega$ = 1100
cm$^{-1}$. The data was reproducible in 8 traces on separate days.
As T decreases from 300K, Re($\theta_{F}$) (and Im($\theta_{F}$))
increases (decreases) gradually. At T $\leq$ 100K, Re($\theta_{F}$)
decreases with lowering T opposite to its high-T behavior and with a
larger slope. The Im($\theta_{F}$) exhibits a sudden step-like jump
at T=$100$K.  This sudden change in $\theta_{F}$ represents, as we
will discuss, the onset of contributions from coherent
quasiparticles below ~$100$K.

%

\begin{figure}[t]
%
%
%
\vspace*{0.4cm}\centerline{\includegraphics[width=3.2in]{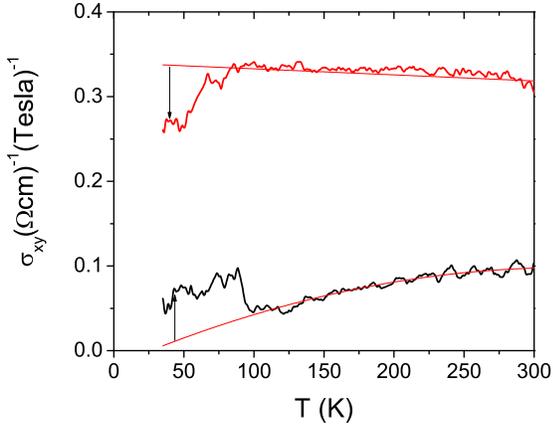}}
\vspace*{0cm} \caption{Infrared Hall conductivity $\sigma_{xy}$.
Dashed lines show polynomial fits of the high temperature part of
data. The arrows represent the low-T deviations from the fit.}
\label{fig:mag}
\end{figure}

The transport functions of interest, the Hall angle,
$\sigma_{xy}(\omega)$, and the Hall coefficient are derived from the
Faraday angle data together with the optical conductivity
$\sigma_{xx}(\omega)$, the film thickness, and substrate index from
the Fresnel analysis of the thin film magneto optics as has been
described elsewhere \cite{Cerne}. $\sigma_{xx}(\omega)$ of single
crystal Na$_x$CoO$_2$ has been reported in the literature from
reflectivity or ellipsometry measurements for different
x's.\cite{wang, hwang, ellipso, lupi} To analyze the magneto optical
data, we employ the $\sigma_{xx}(\omega)$ of x=0.7 \cite{wang}
 and also nearby compositions x=0.75 \cite{hwang} and x=0.82.\cite{ellipso}
The $\sigma_{xx}(\omega)$ spectrum of Na$_{0.75}$CoO$_2$ shows that
the optical conductivity consists of a Drude peak in the Far-IR
range, plus a broad incoherent absorption at mid-infrared
frequencies. These two components show up differently in
$\sigma_{xy}(\omega)$. Near the frequency of our measurement,
$\sigma_{xx}(\omega)$ is only weakly temperature dependent.

We will focus on the Hall conductivity $\sigma_{xy}(\omega)$ because
it is the most fundamental magneto-transport response function. Fig.
2 shows $\sigma_{xy}(\omega)$ derived from $\theta_{F}$  and
$\sigma_{xx}(\omega)$. Re($\sigma_{xy}$)and Im($\sigma_{xy}$) show
similar T-dependence as $\theta_{F}$ because $\sigma_{xx}(\omega)$
is practically featureless and depends only weakly on temperature.
To understand the interesting low-T behavior, we first fit the
high-T data with a simple polynomial in T curve shown as dashed
lines, and assume that they extend to T=0. We label the deviations
of $\sigma_{xy}$ from the fit as $\Delta \sigma_{xy}(\omega)$=
Re($\Delta \sigma_{xy}) +i \cdot$ Im($\Delta \sigma_{xy}$).  We
attribute $\Delta \sigma_{xy}(\omega)$ to the coherent response of
quasiparticles in Na$_{0.7}$CoO$_2$ which we model using a Drude
conductivity form,

\begin{eqnarray}
\Delta\sigma_{xy}(\omega)=\frac{\omega_p^2 \omega_{H}}
{4\pi}\cdot\frac{1}{(\gamma_{H}-\iota\omega )^{2}}
\end{eqnarray}

Here $\omega_p$ and $\omega_{H}$ are the plasma frequency and the
Hall frequency, respectively. $\gamma_{H}$ is the Hall scattering
rate. The former quantities are written as $\omega_p^2=4\pi \frac
{nq^2}{m^{*}}$ and $\omega_{H}=\frac {qB}{m_{H}c}$ where n is the
carrier density, $m^{*}$ and $m_{H}$ are the effective and Hall
mass, respectively.
 We used Eq.(2) to extract $\omega_p^2 \omega_{H}$ and
$\gamma_{H}$ as a function of temperature from the data. These results are displayed in Fig.3.
%
%

\begin{figure}[t]
%
%
%
%
\vspace*{0.4cm}\centerline{\includegraphics[width=3.2in]{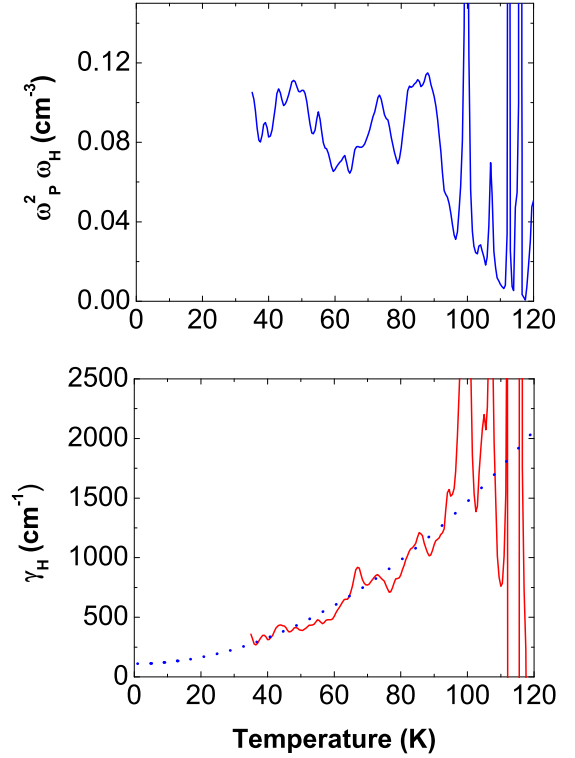}}
\vspace*{0cm} \caption{Drude carrier analysis of the low-T
$\Delta\sigma_{xy}$. (a) Squared plasma frequency
$\omega_{P}^{2}(T)$ times Hall frequency $\omega_{H}(T)$ (b) Hall
scattering rate $\gamma_{H} (T)$. Dashed line shows a fit with
$\gamma_{H}(T)$=$\gamma_{0}$ + a$T^{2}$. } \label{fig:opt}
\end{figure}

$\omega_{P}^{2} \omega_{H}(T)$ is seen to decrease as T increases
and vanishes by T$\sim $120K. Positive $\omega_{H}$ corresponds to
hole conduction, $q>0$. $\gamma_{H}(T)$ is observed to increase
continuously with T. Close to 120 K, the data become increasingly
noisy because $\Delta \sigma_{xy}$ is small in this T-range, and it
enters into denominator in calculating $\omega_{P}^{2}$ and
$\gamma_{H}(T)$. Angle resolved photoemission spectroscopy (ARPES)
of Na$_{0.7}$CoO$_{2}$ by Hasan et al \cite{Hasan04} showed that
Fermi surface consists of a large hole pocket centered at $\Gamma$
point of Brillouin zone. The Fermi surface is nearly isotropic with
k$_F$=0.65{\AA}$^{-1}$. The Fermi surface volume corresponds to
carrier density n$\cong$8.4$\cdot$10$^{21}$ cm$^{-3}$ or 0.32 hole
per Co plane. It is comparable to the nominal hole doping 1-x=0.3.
Using this n in $\omega_p^2 \omega_{H}$ = $\frac {4\pi e^{3}B}{c}
\frac {n}{m^{*}m_{H}}$ at T=30K, and assuming m$^{*}$ = m$_{H}$ we
find m$^{*}$=9.3. The large m$^{*}$ shows strong renormalization of
quasiparticle band, as seen by a weak dispersion of quasiparticle
band in the ARPES data which also is reflected in
 electronic specific heat measurements.\cite{Ando,
Bruhwiler} From n, m$^{*}$, and $\gamma$ , we estimate dc
resistivity $\rho$ =$\frac {m^{*} \gamma}{ne^2}$= 1.7$\cdot$
10$^{-4}$ $\Omega$ cm. This value is comparable with measured
dc-value, 0.1 m$\Omega$ cm. If the suppression were due to a
divergence of the effective mass $m_{H}\rightarrow \infty$ as
T$\rightarrow$ 120K, the electronic specific heat would show a
diverging behavior at this temperature in contrast to the reported
experimental data.\cite{C300K} It is therefore more likely that the
observed behavior comes from a reduction of coherent carrier
density. The suppression of the quasiparticle peak was also seen in
ARPES measurement. \cite{Hasan04}

We observe that $\gamma_{H} (T)$ increases with T approximately as
T$^{2}$ similar to a Fermi liquid. However, the magnitude of the
scattering is large, $\gamma_{H}\gg$ T in contrast to the cuprates.
Also the strong temperature dependence in the mid IR is in contrast
with the observed behavior of the scattering rate deduced from
$\sigma_{xx}(\omega)$ in Na$_{0.65}$CoO$_{2}$. \cite{hwang} At
frequencies around 1000 cm$^{-1}$ they find $\gamma_{xx}$ to be
independent of temperature for T $<$100 K and only weakly
temperature dependent for T $>$100 K. However, in their analysis
they are assuming a one component model of conduction which we
believe our IR Hall data refutes.  The strong temperature dependence
of $\gamma_{H}$ in the mid IR is also not the expected behavior for
a Fermi liquid.  In Fermi liquids, the effective scattering rate for
$\sigma_{xx}(\omega)$ is generally observed to behave as $\gamma
\propto (\omega^2 + (\pi kT)^2)$. At 120K, the first term is ~13
times the second term at our frequencies so that only a weak
T-dependence would be expected..  This implies a weak $\omega$
dependence of $\gamma_{H}$ in $\sigma_{xy}(\omega)$. Such anomalous
behavior of $\gamma_{H}$ has also been observed the optimally doped
cuprates by IR Hall measurement and interpreted by
Kotani.\cite{htsc2} In this theory, the suppression of the frequency
dependence of $\gamma_{H}$ is a consequence of current vertex
corrections $\sigma_{xy}(\omega)$ for systems close to a spin
density wave antiferromagnetic transition. Na$_{0.7}$CoO$_2$ is
known to undergo a SDW transition at T=20K.\cite{sdw}

The high-T part of $\sigma_{xy}(\omega)$ is associated with
incoherent carriers. Hwang \emph{et al} found that \cite{hwang}
$\sigma_{xx}(\omega)$ consists of the Drude component and a broad
mid-infrared incoherent peak. At low T, the latter has a localized
line shape centered  at $\sim$1000 cm $^{-1}$. If we model it with a
Lorentzian oscillator , its contribution to Hall conductivity
$\sigma_{xy}(\omega)$ can be written as
\begin{eqnarray}
\sigma_{xy}(\omega)=\frac{\Omega_p^2
\Omega_{H}}{4\pi}\cdot\frac{\omega^2}
{(\omega^2-\Omega_0^2+i\omega\Gamma)^{2}}
\end{eqnarray}

where $\Omega_p^2$, $\Omega_0$, and $\Gamma$ are the strength,
center frequency, and width of the oscillator. $\Omega_{H}$
represent the the Hall frequency of the incoherent carriers. From
$\sigma_{xx}(\omega)$ data, we adopt $\Omega_0$=1000 cm$^{-1}$,
$\Gamma$=5000cm$^{-1}$ and $\Omega_p^2$ similar to that of the Drude
$\omega_p^2$. Since $\Omega_0$ almost coincides with $\omega$ (=1100
cm$^{-1}$) of our experiment, we have Im($\sigma_{xy}$)$\cong$ 0.
This is consistent with our data extrapolated at T=30K. The positive
$\sigma_{xy}$ shows hole character and suggests that some of the
doped holes become incoherent due to some localization process such
as electron-phonon or electron-spin interaction. Such localization
may explain the Curie-Weiss type magnetic susceptibility coexisting
with the metallicity in Na$_{0.7}$CoO$_2$.\cite{Ong} In our model,
$\Omega_0$ is the most interesting parameter. If $\Omega_0$ is
assumed to decrease from 1000 cm$^{-1}$, Re($\sigma_{xy}$) decreases
and Im($\sigma_{xy}$) increases, as the Hall conductivity data
behaves with increasing T. $\sigma_{xy}(\omega)$ at T=300K are
produced if we take $\Omega_0$ = 100 cm$^{-1}$. This suggests that
the incoherent peak shifts to lower frequency with temperature. The
opposite T-dependences of Re($\sigma_{xy}$) and Im($\sigma_{xy}$) ,
increasing and decreasing respectively with T, are produced only by
the $\Omega_0$ shift but not by other parameters of Eq.(2). In
$\sigma_{xx}(\omega)$ the spectral weight is transferred from Drude
part to higher frequency, $\omega <$ 1000 cm$^{-1}$ with increasing
T.\cite{hwang} The incoherent peak as a result gains spectral
weight. Also the peak center effectively appears to shift to lower
$\omega$. At T=300K, $\sigma_{xx}(\omega)$ shows a localized line
shape centered at FIR range ~200 cm$^{-1}$.

%

\begin{figure}[t]
%
%
%
%
\vspace*{0.4cm}\centerline{\includegraphics[width=3.8in,,angle=0]{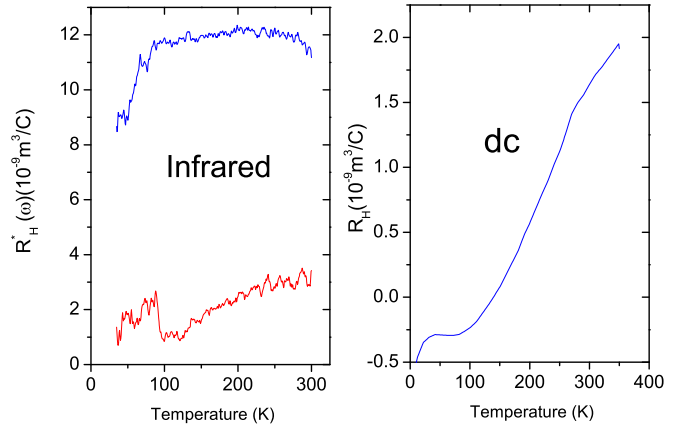}}
\vspace*{0cm} \caption{Infrared Hall coefficient R$^{*}_{H}(\omega)$
and dc-Hall coefficient R$_{H}$.} \label{fig:opt}
\end{figure}

The Dc-Hall effect of Na$_{0.7}$CoO$_2$ is also anomalous: The Hall
coefficient R$_{H}$ is negative at low-T. It becomes positive at T
$>$ 200K and then increases linearly with T without saturation up to
as high as T=500K.\cite{Foo} In a non-interacting model Holstein
showed that the Hall mobility on a triangular lattice is anomalously
enhanced.\cite{Holstein} Shastry $\emph{et al}.$ have studied an
interacting case in the tJ model and found that the high frequency
limit Hall coefficient R$^{*}_{H}(\omega)$ has a T-linear behavior
similar to the measured DC-R$_{H}$. \cite{Shastry2,Koshibae}
However, how this high frequency result relates to the DC R$_{H}$ is
not yet established. It is within this context, that we have studied
the ac-Hall effect and compare it with the dc-Hall result. In Fig.4
we show the infrared Hall coefficient R$^{*}_{H}$ obtained from
R$^{*}_{H} (\omega) =
\frac{\sigma_{xy}(\omega)}{\sigma^2_{xx}(\omega)}$ and compare it
with the measured dc-R$_{H}$ taken on the same film.

Note that the real part of R$^{*}_{H}(\omega)$ is nearly independent
of T for T$>$ 100K in contrast to the dc result. Also there is no
sign change. The imaginary part is much weaker than the real part.
In strongly correlated electron systems, Shastry and Shraiman showed
that the ac-R$^{*}_{H}(\omega)$ simplifies to the the familiar
semiclassical relation R$^{*}_{H}$ = $\frac{1}{nqc}$, ($n$ and $q$
being the carrier density and charge).\cite{Shastry1, Koshibae} The
relation is considered to become valid when the frequency is much
higher than the carrier hopping energy t, ($\omega \gg t$) which
applies to our case, $\omega$ = 1100$cm^{-1}$, t=11 meV.
\cite{Hasan04} The R$^{*}_{H}(\omega)$ then appears to show the
constant density of hole carrier at T$>$ 100K. The Hall constant at
low frequency and dc-limit can be dominated by the self energy due
to strong electron correlation. The non-simple behavior of
dc-R$_{H}$ may represents such effect. Clearly the infrared Hall
effect provides a good starting reference to understand the
anomalous behavior of dc-R$_{H}$. At T$<$ 100K,
$\sigma_{xy}(\omega)$ have contributions from the quasiparticles and
also the incoherent carriers so that R$^{*}_{H}(\omega)$ will be
rather complicated.

In a heavy fermion(HF) metal such as CePd$_{3}$ and UPt$_{3}$, the
condution carrier is hybridized with the local magnetic moment and
leads a narrow low-frequency Drude conductivity. As T increases, the
coherence is disturbed by thermal fluctuation. Then the Drude weight
decreases and disappears at the Kondo temperature T$_{K}$.\cite{RMP}
In Na$_{0.7}$CoO$_2$, the holes coexist with the local Co magnetic
moment of the Curie-Weiss like susceptibility. The effective mass
m$^{*}$ ($\cong$10) is fairly large. Thus it is tempting to consider
the Drude peak and the $\omega_p^2$ suppression to stem from a
similar origin. However, T=100K is an order higher than typical HF
which suggests a strong hybridization, while m$^{*}$=10 , which is
small in the HF standard, is inconsistent with that. Further study
is needed to understand the unusual behavior of the Drude
conductivity.

We thank J.S. Ahn for his help on sample preparation. This work was
supported by the KRF Grant No. 2005-015-C00137, by the KRF Grant No.
2005-070-C00044. We also acknowledge financial support by CSCMR
(Center for Strongly Correlated Material) through KOSEF and BK21
core program at the Dept. of Physics, University of Seoul, Korea.
The work at University of Maryland was supported by NSF grant
DMR-0303112. JHC was supported by Grant No. KRF-2006-005-J02801 from
the Korea Research Foundation

%

%
%
%
%

%
%
%


\end{document}